\documentclass[useAMS,usenatbib]{mn2e}

\usepackage[psamsfonts]{amssymb}
\usepackage[dvips]{graphicx}
\usepackage{amsmath,alltt}
\usepackage{multirow}
\usepackage{rotating}

\title[Dissecting the Colour-Magnitude Diagram]{Dissecting the
  Colour-Magnitude Diagram:  A Homogeneous Catalogue of Stellar Populations in
  Globular Clusters}
\author[Nathan Leigh, Alison Sills and Christian Knigge]{Nathan Leigh$^{1}$,
  Alison Sills$^{1}$, Christian Knigge$^{2}$\thanks{E-mail:
    leighn@mcmaster.ca (NL);
    asills@mcmaster.ca (AS); christian@astro.soton.ac.uk (CK)} \\
$^{1}$Department of Physics and Astronomy, McMaster University,
1280 Main St. W., Hamilton, ON, L8S 4M1, Canada \\
$^{2}$School of Physics and Astronomy, University of Southampton,
Highfield, Southampton, SO17 1BJ, United Kingdom}
\begin{document}

\pagerange{\pageref{firstpage}--\pageref{lastpage}} \pubyear{2010}

\maketitle

\label{firstpage}

\begin{abstract}
We present a homogeneous catalogue for blue straggler, red giant
branch, horizontal branch and main-sequence turn-off 
stars in a sample of 35 clusters taken from the ACS Survey
for Globular Clusters.  As a result of the superior photometry and 
relatively large field of view offered by the ACS data, this new 
catalogue is a significant 
improvement upon the one presented in \citet{leigh07}.  Using our 
catalogue, we study and compare the radial distributions of the
different stellar populations.

We have confirmed our previous result \citep{knigge09} that there is a
clear, but sub-linear, correlation between the number of blue
stragglers found in the cluster core and the total stellar mass
contained within it.  By considering a larger spatial extent than just
the core, our results suggest that 
mass segregation is not the dominant effect contributing to the observed 
sub-linearity.  We also investigate the radial distributions of the
different stellar populations in our sample of clusters.  Our results
are consistent with a linear 
relationship between the number of stars in these populations and the
total mass enclosed within the same radius.  Therefore, we conclude
that the cluster dynamics does not significantly affect the relative 
distributions of these populations in our sample.
\end{abstract}


\section{Introduction} \label{intro}
Colour-magnitude diagrams (CMDs) are one of the most important tools
available to astronomers for studying stellar evolution, stellar
populations and star clusters.  And yet, there remain several
features found in CMDs whose origins are still a mystery.
Examples include horizontal branch (HB) morphology, the presence of
extended horizontal branch (EHB) stars, and blue stragglers (BSs) 
\citep[e.g.][]{sandage53, zinn96,
  peterson03, dotter10}.  Previous studies have shown that the
observed differences in the HBs of Milky Way globular clusters (GCs)
are related to metallicity \citep{sandage60}, however at least one
additional parameter is required to explain the spread in their
colours.  Many cluster properties have been suggested as possible
Second and Third Parameters, including age, central density and
cluster luminosity, although no definitive candidates have been
identified \citep[e.g.][]{rood73, fusi93}.  An 
explanation to account for the existence of BSs has proved equally
elusive.  Many BS formation mechanisms have been proposed,
including stellar collisions \citep[e.g.][]{leonard89, sills99} and
binary mass-transfer \citep{mccrea64, mathieu09}.  However, no clear
evidence has yet emerged in favour of a dominant formation
mechanism.    

In short, we still do 
not understand how many of the physical processes operating within
star clusters should affect the appearance of CMDs
\citep[e.g.][]{fusi92, buonanno97, ferraro99, beccari06}.  In general,
the importance of these processes can be constrained by looking for
correlations between particular features in CMDs and cluster properties
that serve as proxies for different effects.
For example, the central density can be used as a rough proxy for the
frequency with which close dynamical encounters occur.  Similarly, the
cluster mass can be used as a proxy for the rate of two-body relaxation.  
Once the relevant effects are accounted for, CMDs can continue to
provide an ideal tool to further our 
understanding of stellar evolution, stellar populations and star
clusters. 

It is now clear that an important interplay
occurs in clusters between stellar dynamics and stellar evolution.
For example, dynamical models have shown that 
star clusters expand in response to mass-loss driven by stellar
evolution, particularly during their early evolutionary phases when 
massive stars are still present \citep[e.g.][]{chernoff90, portegieszwart98,
  gieles10}.  Mass-loss resulting from stellar evolution has also been
proposed to cause horizontal branch stars to exhibit more
extended radial distributions relative to 
red giant branch and main-sequence turn-off (MSTO) stars in
globular clusters having short 
central relaxation times relative to the average HB lifetime
\citep[e.g.][]{sigurdsson95, leigh09}.  This can be understood as
follows.  Red giant branch (RGB) stars
should be more mass segregated than other stellar populations since
they are among the most massive stars in GCs.  HB stars, on the other
hand, are much less massive since RGB stars experience significant
mass loss upon evolving into HB stars.  Consequently, two-body
relaxation should act to 
re-distribute HB stars to wider orbits within the cluster potential
relative to RGB and MSTO stars \citep{spitzer75}, provided the average
HB lifetime is shorter than 
the central relaxation time.  Studies have shown that the radial
distributions of the HB populations in some GCs could differ from
those of other stellar populations.  For
instance, \citet{saviane98} 
presented evidence that blue HB stars could be more centrally
concentrated than red HB and sub-giant branch stars in the GC NGC
1851.  Conversely, \citet{cohen97} showed that blue HB stars could be
centrally depleted relative to other stellar types in the GC NGC
6205.  To date, no clear evidence has been found linking the spatial
distributions of HB stars to any global cluster properties.  

Peculiar trends have also been reported for the radial distributions
of RGB stars.  
For example, a deficiency of bright red giants has been observed 
in the GC NGC 1851 \citep[e.g.][]{iannicola09}.  \citet{sandquist07} 
discussed the possibility that this deficiency could be the result of
strong mass loss on the RGB.  
Alternatively, some authors have suggested that dynamical effects could
deplete red giants.  For instance, 
giants could experience collisions more frequently than other stellar
populations due to their larger cross-sections for collision
\citep{beers04}.

One important example of the interplay that occurs in clusters between
stellar evolution and stellar dynamics can be found in the study
of blue stragglers.  Found commonly in both open and globular clusters, BSs are
thought to be produced by the addition of hydrogen to the cores of
low-mass main-sequence (MS) stars, and therefore appear as an
extension of the MSTO in cluster CMDs \citep{sandage53}.  This can
occur via multiple channels, most of which involve the mergers of
low-mass MS stars since a significant amount of mass is typically 
required to reproduce the observed locations of BSs in CMDs
\citep[e.g.][]{sills99}.  Stars in close binaries can merge if enough
orbital angular momentum is lost, which can be mediated by dynamical
interactions with other stars, magnetized stellar winds, tidal
dissipation or even an outer triple companion
\citep[e.g.][]{leonard92, li06, perets09, dervisoglu10}.
Alternatively, MS stars can collide directly, although this is 
also thought to usually be mediated by multiple star systems
\citep[e.g.][]{leonard89, fregeau04, leigh10}.  Finally, BSs have also
been hypothesized to form by 
mass-transfer from an evolving primary onto a normal MS companion
via Roche lobe overflow \citep{mccrea64}.  

Whatever the dominant BS
formation mechanism(s) operating in dense star clusters, it is now
thought to somehow involve multiple star systems.  This was shown 
to be the case in even the dense cores of GCs \citep{knigge09} where
collisions between single stars are thought to occur frequently
\citep{leonard89}.  In \citet{knigge09}, we showed that the numbers of
BSs in the cores 
of a large sample of GCs correlate with the core masses.  We 
argued that our results are consistent with what is expected if BSs
are descended from binary stars.  \citet{mathieu09} also showed
that at least $76\%$ of the BSs in the old open cluster NGC 188 have 
binary companions.  Although the nature of these companions remains
unknown, it is clear that binaries played a role in the
formation of these BSs.  Dynamical
interactions occur frequently enough in dense clusters that they
are also expected to be at least partly responsible for the observed
properties of BSs.  It follows
that the current properties of BS populations
should reflect the dynamical histories of their host clusters.  As a
result, BSs could provide an indirect means of probing
the physical processes that drive star cluster evolution.

In this paper, we present a homogeneous catalogue for red giant
branch, main-sequence turn-off, horizontal branch and blue
straggler stars in a sample of 35 Milky Way (MW) GCs
taken from the Advanced Camera for Surveys
(ACS) Survey for Globular Clusters \citep{sarajedini07}.  With this
catalogue, we investigate two important issues related to stellar
populations in GCs.  First, we test the observational correlation
found for BSs presented in \citet{knigge09} by re-doing the study with
newer and more accurate photometry.  The larger spatial coverage
considered in our new sample 
offers an important additional constraint for the origin of this correlation.  
Second, we perform the same statistical comparison for RGB, HB and
MSTO stars in 
order to study their radial distributions.  This will allow us to test
some of the results and hypotheses introduced in \citet{leigh09},
where we first presented this
technique for studying stellar populations.  In
particular, we found evidence for a surplus of RGB stars in
low-mass GC cores relative to MSTO stars.  However,
we concluded that the study needed to be re-done with better
photometry.  The ACS data are of sufficiently high quality to address
this issue.

In Section~\ref{method}, we present our selection criteria to
determine the numbers of BS, RGB, HB and MSTO stars located in the central
cluster regions.  The spatial coverage of the photometry extends out
to several core radii from the cluster centre for most of the clusters
in our sample.  For these clusters, we have obtained number counts within
several circles centred on the cluster centres provided in
\citet{goldsbury10} 
for various multiples of the core radius.  This catalogue is presented in
Section~\ref{results}.  In this section, we also present a comparison
between the sizes of the different stellar populations and the total
stellar masses contained within each circle and annulus.  Finally,
we discuss our results for both BSs and the other 
stellar populations in Section~\ref{discussion}.   

\section{Method} \label{method}

In this section, we present our sample of CMDs and define our
selection criteria for each of the different stellar populations.  We
also discuss the spatial coverage offered by the ACS sample, and
describe how we obtain estimates for several different fractions of
the total cluster mass from King models.

\subsection{The Data} \label{data}

The data used in this study consists of a sample of 35 MW GCs taken
from the ACS Survey for Globular Clusters \citep{sarajedini07}.\footnote[1]{The
data can be found at http://www.astro.ufl.edu/~ata/public\_hstgc/.}  The
ACS Survey provides unpecedented deep photometry in the F606W ($\sim$
V) and F814W ($\sim$ I) filters 
that is nearly complete down to $\sim 0.2$ M$_{\odot}$.  In other
words, the 
CMDs extend reliably from the HB all the way down to about 7
magnitudes below the MSTO.  We have confirmed that
the photometry is nearly complete above at least 0.5
magnitudes below the MSTO for every cluster in our sample.  This was
done using the results of artificial star tests taken from  
\citet{anderson08}, and confirms that the photometric quality of
the stellar population catalogue presented in this paper is very
high.\footnote[2]{Artificial star tests were obtained directly from Ata
Sarajedini via private communication.}  We have also considered
foreground contamination by field stars, and it is negligible.

Each cluster was centred in the ACS field, which
extends out to several core radii from the cluster 
centre in most clusters and, in a few cases, beyond even 15 core
radii.  Coordinates for the cluster centres were taken from 
\citet{goldsbury10}.  These authors found their centres by fitting
a series of ellipses to the density distributions within the inner 2'
of the cluster centre, and computing an average value.  The core
radii were taken from \citet{harris96}.

\subsection{Stellar Population Selection Criteria} \label{criteria}

In order to select the number of stars belonging to each stellar
population, we define a series of lines in the (F606W-F814W)-F814W
plane that act as boundaries enclosing each of the different stellar
populations.  
To do this, we fit theoretical isochrones taken from \citet{dotter07}
to the CMDs of every cluster in our sample.  Each isochrone
was generated using the metallicity and age of the cluster, and fit to
its CMD using the corresponding distance modulus and extinction
provided in \citet{dotter10}.  
The MSTO was then defined using our isochrone fits by selecting the
bluest point along the MS.  This acts as our primary point of
reference for defining the boundaries in the CMD for the different
stellar populations.  Consequently, the selection criteria
provided in this paper are a significant improvement upon the criteria
presented in \citet{leigh07}, and our new catalogue for the
different stellar populations is highly homogeneous.

Two additional points of reference must also
be defined in order for our selection criteria to be applied 
consistently from cluster-to-cluster.  First, the selection criteria 
for the HB are determined by fitting a line through the approximate
mid-point of the points that populate it in the CMD.  This
line is then used to define upper and lower boundaries for the
HB.  Theoretical isochrones become highly uncertain at the HB, so it
is necessary to specify this additional criterion by eye.  Second, the
lower boundary of the RGB is defined for each cluster as 
the point along its isochrone corresponding to a helium core mass of
0.08 M$_{\odot}$.  We do not include RGB
stars brighter than the HB since the tilt 
of the upper RGB varies significantly from cluster-to-cluster, 
presenting a considerable challenge for the consistency of our
selection criteria.  Moreover, the distinction between RGB and
asymptotic giant branch stars in the CMD is often ambiguous.

\begin{figure}
\begin{center}
\includegraphics[width=\columnwidth]{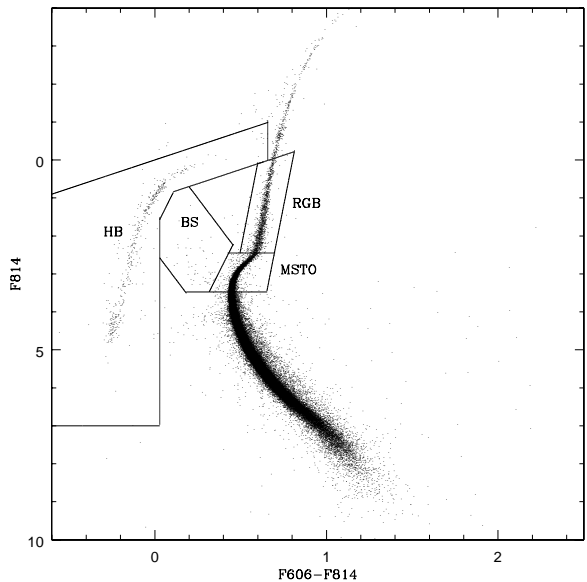}
\end{center}
\caption[Plot showing the parameter space in the (F606W-F814W)-F814W
plane defining each of the different stellar populations for the GC
NGC 6205]{Colour-magnitude diagram for the Milky Way globular cluster
  NGC 6205.  Boundaries enclosing the parameter space in the 
  (F606W-F814W)-F814W plane that define each of the different stellar
  populations are indicated with solid lines, as dedscribed in the
  text.  Absolute 
  magnitudes are shown, converted from apparent magnitudes using the
  distance modulii and extinctions provided in \citet{dotter10}.
  Labels for blue straggler, red giant branch, horizontal branch and
  main-sequence turn-off stars are indicated.  Stars with large
  photometric errors have been omitted from this plot.
\label{fig:fig1}}
\end{figure}

Example selection criteria for each of the different stellar
populations are shown in Figure~\ref{fig:fig1}.  Formal definitions
for the boundaries in the cluster CMD 
that define the BS, RGB, HB and MSTO populations are provided in
Appendix~\ref{appendix}.  We note that the sizes of our selection
boxes have been chosen to accomodate the photometric errors, which
contribute to broadening the various evolutionary sequences in the
CMD.  With superior photometry, the sizes of our selection
boxes could therefore be reduced.  This would further 
decrease contamination from field stars in our samples.

\subsection{Spatial Coverage} \label{spatial}

\begin{figure}
\begin{center}
\includegraphics[width=\columnwidth]{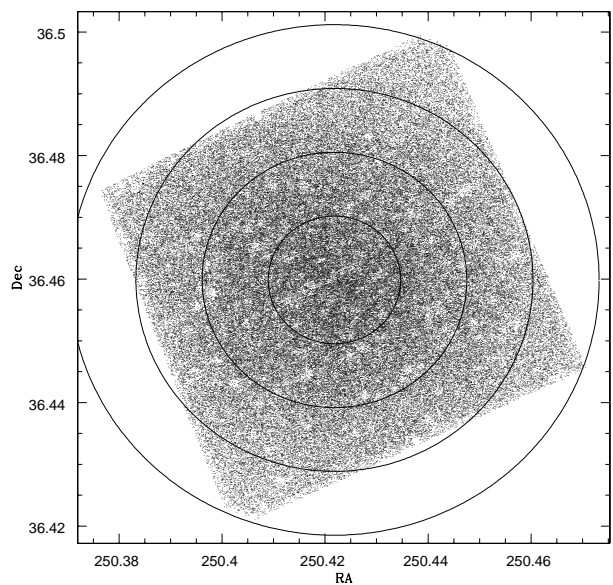}
\end{center}
\caption[Plot showing the RA and Dec coordinates for all stars in NGC
6205]{RA and Dec coordinates for all stars in the GC NGC 6205.
  Circles corresponding to one, two, three and four core radii are 
  shown.  
\label{fig:fig2}}
\end{figure}

The ACS field of view extends out to several core radii from the
cluster centre for nearly every cluster in our sample.  Consequently,
we have obtained estimates for the number 
of stars contained within four different circles centred on the central
cluster coordinates provided in \citet{goldsbury10}.  We list
these numbers only for clusters for which the indicated circle is
completely sampled by the field of view.  
The radii of the circles were taken to be integer multiples of the
core radius, and we focus our attention on the inner four core radii
since the field of view extends beyond this for only a handful of the
clusters in our sample.  An example
of this is shown in Figure~\ref{fig:fig2}.  The numbers we
list are cumulative, so that entries for each circle include
stars contained within all smaller circles.  

\subsection{King Models} \label{king}

In order to obtain accurate estimates for the total stellar mass
contained within each circle, we generated 
single-mass King models calculated using the method of
\citet{sigurdsson95} to obtain luminosity density profiles
for the 
majority of the clusters in our sample.  The profiles were obtained using
the concentration parameters of \citet{mclaughlin05} and the central
luminosity densities of \citet{harris96} for each cluster in
\citet{mclaughlin05} that overlaps with our sample.  We then
integrated the derived 
luminosity density profiles numerically in order to estimate the total
stellar light contained within each circle.  After removing clusters
with high concentration parameters \citep{harris96} for which King
models are known to provide a poor fit, we multiplied the total
stellar light by a mass-to-light ratio of 2 in order to obtain
estimates for the total stellar mass contained within each circle.
Calculating the total stellar mass contained within each
circle from King models requires a number of assumptions that we will
discuss fully in Section~\ref{discussion}.

\section{Results} \label{results}

In this section, we present our catalogue along with the results of
our comparisons between the 
sizes of the different stellar populations and the total stellar mass
contained within each circle and annulus.

\subsection{Catalogue}  \label{catalogue}

The numbers of BS, RGB, HB and MSTO stars found within several
different circles are shown for all clusters in
Table~\ref{table:catalogue}, along with the total number of stars 
with magnitudes brighter than 0.5 mag below the MSTO.  Number counts
are only shown whenever the spatial coverage is complete within the
indicated circle.

\subsection{Population Statistics} \label{statistics}

How can we use our catalogue to learn which, if any, 
cluster properties affect the appearance of CMDs?
One way to accomplish this is by plotting the size of a
given stellar population in a particular circle versus the total stellar
mass contained within it.  From this, lines of best fit can be found
that provide equations relating the size of each stellar population to 
the total stellar mass contained within each circle.  As described
below, this is ideal for probing the effects of the cluster dynamics
on the appearance of CMDs.  

The rate of two-body relaxation for a cluster can be approximated
using the half-mass relaxation time \citep{spitzer87}:
\begin{equation}
\label{eqn:t-rh}
t_{rh} = 1.7 \times 10^5[r_h(pc)]^{3/2}N^{1/2}[m/M_{\odot}]^{-1/2} years,
\end{equation}
where $r_h$ is the half-mass radius, $N$ is the total number of stars
within $r_h$ and $m$ is the average stellar mass.  
The half-mass radii of MW GCs are remarkably similar independent of mass,
and simulations have shown that $r_h$ changes by a factor of at most a few
over the course of a cluster's lifetime \citep{murray09, henon73}.  
The GCs that comprise our sample show a range of masses spanning roughly 3
orders of magnitude, and have comparably old ages \citep{deangeli05}.
Therefore, Equation~\ref{eqn:t-rh} suggests that the degree of
dynamical evolution (due to two-body relaxation) experienced by a
cluster is primarily determined by the total cluster mass for the 
GCs in our sample.  In particular, more massive clusters are less
dynamically evolved, and vice versa.  Consequently, if the size of a
given stellar population is affected by two-body relaxation, the
effects should be the most pronounced in the least massive clusters in
our sample.  
Additionally, the rate of (direct) stellar collisions increases 
with increasing cluster mass \citep[e.g.][]{davies04}.  This suggests
that, if a given stellar population is affected by collisions, the
effects should be the most pronounced in the most massive clusters in
our sample.  Therefore, by comparing the size of each stellar population
to the total stellar mass contained within a given circle, the 
effects of the cluster dynamics can be quantified.  
This technique also ensures a normalized and consistent comparison
since it accounts for cluster-to-cluster
differences in the fractional area sampled by the ACS field of view.
That is, we are consistently comparing the same
structural area for each cluster.  The validity and implications of
all of these assumptions will be discussed further in
Section~\ref{discussion}. 

\begin{figure}
\begin{center}
\includegraphics[width=\columnwidth]{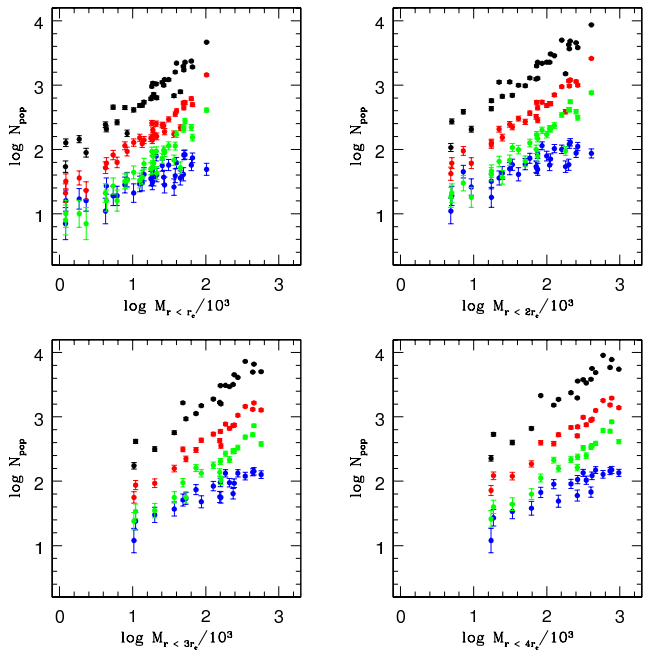}
\end{center}
\caption[Plot showing for each circle the logarithm of the number of
stars belonging to each stellar population as a function of the
logarithm of the total stellar mass]{The logarithm of the number of
  stars belonging to each stellar population is shown for each circle
  as a function of the logarithm of the total 
stellar mass.  From left to right and top to bottom, each frame
corresponds to number counts contained within a circle having a
radius of $r_c$, $2r_c$, $3r_c$ and $4r_c$.  Blue corresponds to blue
stragglers, red to red giant branch stars, green to horizontal branch
stars and black to main-sequence turn-off stars.
Estimates for the total
stellar mass contained within each circle were found using single-mass King
models, as described in Section~\ref{king}.  
\label{fig:Mshell_vs_Npop_2x2_cum}}
\end{figure}

\begin{figure}
\begin{center}
\includegraphics[width=\columnwidth]{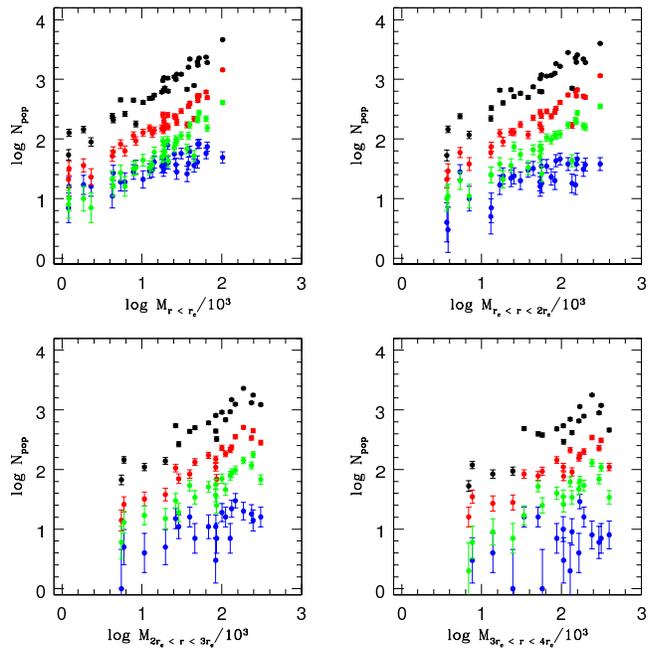}
\end{center}
\caption[Plot showing for each annulus the logarithm of the number of
stars belonging to each stellar population as a function of the
logarithm of the total stellar mass]{The logarithm of the number of
  stars belonging to each stellar population is shown for each annulus
  as a function of the logarithm of the total
stellar mass.  The annulus and colour corresponding to each inset and
stellar population, respectively, are the same as in
Figure~\ref{fig:Mshell_vs_Npop_2x2_cum}.
\label{fig:Mshell_vs_Npop_2x2_noncum}}
\end{figure}

Plots showing the number of stars belonging to each stellar population
as a function of the total stellar mass contained within each circle
are shown in Figure~\ref{fig:Mshell_vs_Npop_2x2_cum}.  Uncertainties
for the number of stars belonging to each stellar population were
calculated using Poisson statistics.  We also plot 
in Figure~\ref{fig:Mshell_vs_Npop_2x2_noncum} the number of stars
belonging to each stellar population as a 
function of the total stellar mass contained in each annulus outside
the core.  That is, we
considered the populations for each annulus individually, as opposed
to considering every star with a 
distance from the cluster centre smaller than the radius of the
outer-most circles.  Recall that we have neglected 
clusters for which
our theoretical King models provide a poor description of the true
density distributions.  This was the case for clusters
in our sample having a high concentration parameter, most of
which are labelled as post-core collapse in \citet{harris96}.

%
%

We performed a weighted least-squares fit for every relation in
Figure~\ref{fig:Mshell_vs_Npop_2x2_cum} and
Figure~\ref{fig:Mshell_vs_Npop_2x2_noncum}.  Slopes and y-intercepts
for these lines are shown in Table~\ref{table:Mshell_vs_Npop_cum} and
Table~\ref{table:Mshell_vs_Npop_noncum}, respectively.  Uncertainties for the
slopes and y-intercepts were 
found using a bootstrap methodology in which we generated 1,000 fake
data sets by randomly sampling (with replacement) number counts from the
observations.  We obtained lines of best fit for each fake data set, fit a
Gaussian to the subsequent distribution and extracted its standard
deviation.  

As shown in Table~\ref{table:Mshell_vs_Npop_cum}, the power-law index
is sub-linear for BSs within the core at much better than the
$3-\sigma$ confidence level, and it is consistent with the slope 
obtained in our earlier analysis presented in \citet{knigge09}.  The
slopes for the BSs are also sub-linear 
at better than the $3-\sigma$ confidence level for all circles outside the
core.  This is also the case for all annuli outside the core, as shown
in Table~\ref{table:Mshell_vs_Npop_noncum}.  Note, however, that the
uncertainties for the BS slopes are very 
large for all annuli outside the core, whereas this is not always the case
for corresponding circles outside the core.  This is the result of the
fact that the number of BSs drops off rapidly outside
the core in several clusters so that the corresponding Poisson
uncertainties, which are given by the square-root of the number of
BSs, are significant.  The rapid decline of BS numbers with 
increasing distance from the cluster centre in these clusters has also
contributed to an 
increased degree of scatter in the relations for annuli outside the core
relative to the corresponding relations for circles outside the core.

The slopes are consistent with being linear for all other stellar
populations in the core within their respective $3-\sigma$ confidence
intervals.  This agrees with the results of our earlier analysis
presented in \citet{leigh09} when we performed the comparison using
the total core masses.  The slopes are also consistent with being 
linear for all circles outside 
the core for both HB and MSTO stars.  The power-law indices are sub-linear 
at the $3-\sigma$ confidence level only for RGB stars, and this is
only the case for circles outside the core.  The power-law index is
nearly unity for the core RGB population, 
yet the associated uncertainty is very large.  Upon closer inspection,
the distribution of power-law indices obtained from our bootstrap
analysis for RGB stars in the core is strongly bi-modal, with
comparably-sized peaks centred at $\sim 0.82$ and $\sim 1.0$.  This
bi-modality is most likely an artifact of our bootstrap analysis
caused by a chance 
alignment of data points in the log M$_{core}$-log N$_{RGB}$ plane.  
Upon performing the comparison for only those stars found within 
particular annuli, our results suggest that the slopes are consistent
with being sub-linear for all stellar 
populations at the $3-\sigma$ confidence level in only the annulus
immediately outside the core (i.e. $r_c < r < 2r_c$).  The slopes are
consistent with being linear for RGB, HB and MSTO stars in all other
annuli.

We also tried 
performing the same comparisons using the total number of stars in each
circle and annulus as a proxy for the total stellar mass.  In this
case, the slopes are 
sub-linear for BSs within all circles and annuli at the
$3-\sigma$ confidence level.  Once again, the uncertainties are very
large for all annuli outside the core, whereas this is not the case
for corresponding circles outside the core.  The slopes are consistent
with being linear within the $1-\sigma$ confidence
interval for all other stellar populations in all circles and annuli.
Our results are therefore inconsistent with those presented in
\citet{leigh09} for the core RGB populations, in which we found that
RGB numbers scale sub-linearly with the number of stars in the core at
the $3-\sigma$ confidence level.  We will discuss the implications of
these new results in Section~\ref{discussion}.

\subsection{Blue Stragglers and Single-Single Collisions} \label{collisions}

\begin{figure}
\begin{center}
\includegraphics[width=\columnwidth]{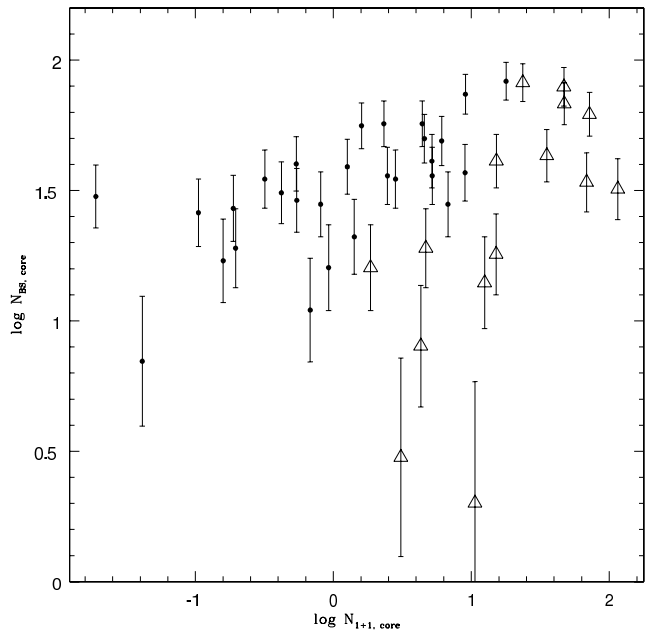}
\end{center}
\caption[Plot showing the logarithm of the number of BSs predicted to
have formed in the core from single-single collisions versus the logarithm of
the observed number of BSs in the core.]{The logarithm of the number
  of BSs predicted to have formed in the core from single-single collisions
  N$_{1+1}$ 
  versus the logarithm of the observed number of BSs in the core N$_{BS}$.
  Filled circles correspond to clusters
having central luminosity densities satisfying log $\rho_0 < 4$, 
whereas open triangles correspond to dense clusters for which log
$\rho_0 > 4$.  The adopted definition for N$_{1+1}$ has been provided
in the text.
\label{fig:Ncoll_vs_Nbs}}
\end{figure}

As a check of our previous results reported in \citet{knigge09}, we
also looked for a correlation between the observed number of BSs in
the cluster core and the number predicted from single-single (1+1)
collisions.  The results of this comparison are shown in
Figure~\ref{fig:Ncoll_vs_Nbs}.  We define the predicted
number of BSs formed from 1+1 collisions as $N_{1+1} =
\tau_{BS}/\tau_{1+1}$, where $\tau_{BS}$ is the average BS lifetime
and $\tau_{1+1}$ is the average time between 1+1 collisions in the
cluster core.  We adopt the same
definition for $\tau_{1+1}$ as used in \citet{knigge09}, and assume
$\tau_{BS} = 1.5$ Gyrs as well as an average stellar mass and radius of
$0.5$ M$_{\odot}$ and $0.5$ R$_{\odot}$, respectively.  We also adopt
a constant mass-to-light ratio of $M/L = 2$ for all clusters.  Central
luminosity densities and velocity dispersions were taken from 
\citet{harris96} and \citet{webbink85}, respectively.

Upon performing a weighted line of best fit for every cluster in our
sample that overlaps with the catalogue of \citet{webbink85}, we find
a power-law index of $0.15 \pm 0.03$ (the uncertainty
was found using the bootstrap methodology described in
Section~\ref{statistics}).  For the subset of dense 
clusters having a central luminosity density satisfying log
$\rho_0 > 4$, we find a power-law index of $0.36 \pm 0.14$.  As before,
we find no significant correlation with collision rate, even for the 
subset of dense clusters.  Although we do find a weak dependence of BS
numbers on collision rate for the entire sample, this is not
unexpected since the collision rate and the core mass are themselves
correlated, and our results suggest that there exists a strong
correlation between BS numbers and the core masses.

\section{Summary \& Discussion} \label{discussion}

In this paper, we have presented a catalogue for BS, RGB,
HB and MSTO stars in a sample of 35 GCs.  Our catalogue provides
number counts for each stellar population within several different
circles centred on the cluster centre.  The radii of the circles were
taken to be integer multiples of the
core radius, and we have focussed on the inner four core radii
since the field of view extends beyond this for only a handful of the
clusters in our sample.  Particular consideration was
given to our selection criteria for the 
different stellar populations in order to ensure that they were
applied consistently from cluster-to-cluster.  In particular, we have
improved upon our previous selection criteria \citep{leigh07} by
fitting theoretical isochrones to the cluster CMDs.  This provides an
unambiguous definition for the location of the MSTO, which acts as the 
primary point of reference for the application of our selection
criteria.  As a result, our new catalogue is highly homogeneous.

We have used our catalogue to quantify the dependence
of the size of each stellar population on the total stellar mass
enclosed within the same radius.  As described in
Section~\ref{statistics}, this provides a means of quantifying the
effects, if any, had by the cluster dynamics in shaping the appearance 
of CMDs above the MSTO.  Below, we summarize the implications of our
results for each of the different stellar populations.
 
\subsection{Blue Stragglers} \label{BSs}

We have confirmed our previous result that the numbers of BSs in the
cores of GCs scale sub-linearly with the core masses
\citep{knigge09}.  That is, we find proportionately larger BS
populations in low-mass GCs.  
There exist several possibilities that could explain the origin of
this sub-linear dependence.  First, we previously suggested 
that this could be an artifact of an anticorrelation between the binary
fraction and the cluster (or core) mass \citep{knigge09}.  This
assertion stems from the fact 
that, if BSs have a binary origin, we expect their numbers to scale
with the core mass as $N_{BS} \sim f_bM_{core}$, where $f_b$ is the binary
fraction in the core.  As before, we find that $N_{BS} \sim
M_{core}^{0.4-0.5}$.  Our result could therefore be explained if $f_b \sim
M_{core}^{-(0.5-0.6)}$.  Second, we suggested in \citet{leigh09} that
the fact that the least massive GCs in our sample should be
more dynamically evolved than their more massive
counterparts could be contributing to the observed sub-linearity for
BSs.  In particular, the low-mass clusters in our sample should have
experienced a significant depletion of their very low-mass stars as a
result of stellar evaporation induced by two-body relaxation 
\citep[e.g.][]{spitzer87, heggie03, demarchi10}.  In turn, this could contribute
to a higher fraction of merger products having masses that exceed that
of the MSTO in low-mass GCs.  As a result, more merger products would 
appear brighter and bluer than the MSTO in these clusters' CMDs,
leading to more merger products being identified as BSs.  Finally,
mass segregation could also be contributing to the observed sub-linear
dependence for BSs.  Again, this is the result of the fact that the
rate of two-body relaxation, and therefore dynamical friction, is in
general the 
fastest in low-mass clusters.  BSs are among the most massive stars in
GCs, so they should rapidly migrate into the core via 
dynamical friction in clusters for which the half-mass relaxation time
is shorter than the average BS lifetime.  It follows that proportionately
more BSs could have drifted into the core via dynamical friction in
low-mass GCs.  This would also contribute to the observed sub-linear
dependence of BS numbers on the core masses. 

This last hypothesis can be tested by comparing our scaling
relations for progressively larger circles outside the core.  If mass
segregation is indeed the cause of the observed sub-linear dependence
of BS numbers on the core masses, then we might expect the power-law
index to systematically increase as we consider progressively larger
circles.  That is, we could be including more BSs that have not yet migrated
into the core via dynamical friction, particularly in the most massive
clusters in our sample.  However, our results suggest 
that the power-law index remains roughly constant for all circles.  
This is the case for both the comparison to the total stellar masses as
well as to the total number of stars contained in each circle.  
The fact that these relations are comparably sub-linear within all
circles can be interpreted as evidence that mass segregation is not
the dominant effect contributing to the 
observed sub-linear dependence of BS numbers on the total stellar
masses (or the total number of stars).  We note that for many of the
clusters in our sample, the spatial coverage is comparable to or
exceeds the half-mass radius.  This is a sufficiently large fraction
of the total cluster area for our comparison to be sensitive to the
effects of mass segregation.  

On the other hand, several GCs are known
to show evidence for a bi-modal BS radial distribution
\citep[e.g.][]{mapelli06, lanzoni07}.  That is, in
these clusters the number of BSs is highest in the central cluster
regions and decreases with increasing distance from the cluster centre
until a second rise in BS numbers occurs in the cluster outskirts.
This seconary outer peak has been shown to occur at a distance 
from the cluster centre that exceeds 20 core radii in several
cases.  Consequently, 
the spatial coverage provided by the ACS data likely does not extend
sufficiently far in most clusters to 
detect any bi-modality in the BS radial distribution.  Nonetheless,
if applied to a 
statistically-significant sample for which the spatial coverage
is complete out to the tidal radius, the technique we have
presented in this paper could provide a powerful constraint for the
origin of the 
bi-modal BS radial distribution observed in several MW GCs by
addressing the role played by mass segregation.

We also tried to correlate the number of BSs observed in 
the cluster core with the number predicted from single-single 
collisions.  As in \citet{knigge09}, we find that BS numbers depend
strongly on
core mass, but not on collision rate.  This also proved to be the case
for the subset of 
dense clusters satisfying log $\rho_0 > 4$.  Our previous
interpretation 
that our results provide strong evidence for a binary, as opposed to a
collisional, origin for BSs in GCs therefore remains the same.

\subsection{Red Giant Branch Stars} \label{RGBs}

The technique used in this paper to compare the sizes of
the different stellar populations was first presented in
\citet{leigh09}.  In that study, we introduced our method 
and applied it to a sample of 56 GCs taken from \citet{piotto02} in
order to study their RGB populations.  Our results 
were consistent with a sub-linear dependence of RGB numbers on the
core masses.  In particular, we found evidence for a surplus of RGB
stars relative to MSTO stars in the cores of low-mass GCs.  We
considered several possible causes for this result, 
but concluded that our analysis should ideally be repeated with
superior photometry in order to properly assess effects such as 
completeness.  Given the high-quality of the ACS data, we are now in
a position to reassess our previous result for RGB stars.

Upon applying our technique to the ACS sample, we find that the
numbers of RGB stars scale linearly with the core masses
to within one standard deviation.  This is also the case for our
comparison to the total number of stars in the core.  This 
suggests that we should reject our previous conclusions for this stellar
population reported in \citet{leigh09}.  Specifically, if we take a
strict $3-\sigma$ limit as our 
criterion for whether or not the slopes are sub-linear at a
statistically significant level, then the core RGB slope reported in
this paper is consistent with being linear whereas this was not the case for
the core RGB slope reported in \citet{leigh09}.  However, if we take a
more stringent criterion for statistical significance,
then there is no inconsistency between our new and old results for RGB
stars, and both slopes are consistent with being linear.  

We also tried comparing the RGB catalogue presented in this paper with
the one presented in \citet{leigh09}.  This showed that the
old RGB numbers are slightly 
deficient relative to the new numbers at the high-mass end.  Although this
difference is not sufficiently large to completely account for the
difference in slopes (for the comparison with the core mass) found
between our new and old RGB catalogues, it works in the right
direction and is likely a contributing factor.  If our uncertainties
are also factored in, then our new and old slopes agree to within one
standard deviation (due mainly to the large uncertainty for the new
slope).  The source of the disagreement between our old and new RGB
catalogues is unclear, and we cannot say whether or not 
incompleteness (in the old data set) is the culprit.  The results of
our artificial star tests have at least confirmed that incompleteness
is not an issue for our new catalogue, however it could certainly have
contributed to the lower RGB numbers reported in \citet{leigh09}.
Indeed, the central cluster density tends to be 
higher in more massive clusters, which should negatively affect
completeness.  It is not clear, however, why this would have affected
RGB stars more than MSTO stars in the old data set.  
Given all of these considerations, we feel that our new results show
that this issue needs to be looked at in more detail before any firm
conclusions can be drawn.  

The evidence in favour of RGB numbers scaling linearly with the core masses 
is interesting.  For one thing, it suggests that two-body relaxation 
does not significantly affect RGB population size relative to other
stellar populations of comparable mass in even the dense central
regions of GCs.  This is not surprising, since two-body 
relaxation is a long-range effect for which the stellar radius plays a
negligible role.   Second, it suggests that collisions do not
significantly deplete RGB stars relative to other stellar populations
despite their much larger radii.  This is because the collision rate
increases with increasing cluster mass, so we would expect RGB stars
to appear preferentially depleted in massive clusters if they are
significantly affected by collisions \citep[e.g.][]{beers04, davies04}.  
Third, it suggests that the sub-linear relation found for BSs does not 
contribute to a sub-linear relation for RGB stars despite the fact that
BSs should eventually evolve to occupy our RGB selection box, as
discussed in \citet{leigh09}.  This is likely the result of the
relatively small sizes of the BS populations in our sample when
compared to the numbers of RGB stars since the rate at which evolved
BSs ascend the RGB is thought to be comparable to the RGB lifetimes of
regular MSTO 
stars \citep{sills09}.  Alternatively, this result could, at least in
part, be explained if
a smaller fraction of BSs end up sufficiently bright and blue to
be identified as BSs in the CMDs of massive GCs.  In other words, it
could be that a larger fraction of BSs are 
hidden along the MS in massive clusters, as discussed in
Section~\ref{BSs}.  In this case, the
contributions to RGB populations from evolved BSs could be comparable
in all clusters, in which case a linear relationship between RGB
numbers and the core masses would be expected.  Finally, evolved BSs
would be expected to have a negligible impact on RGB population size
if the average BS lifetime is considerably longer than the lifetimes
of RGB stars.  This effect is difficult to quantify, however, given
that BS lifetimes are poorly constrained 
in the literature \citep[e.g.][]{sandquist97, sills01}.

\subsection{Horizontal Branch Stars} \label{HBs}

Our results suggest that HB numbers scale linearly with the core
masses.  This can be interpreted as evidence that two-body relaxation
does not significantly affect the radial distributions of HB 
stars in GCs relative to the other stellar populations above the
MSTO.  One reason to perhaps expect that two-body relaxation should
affect the spatial distributions of HB stars stems from the fact 
that RGB stars are among the most massive stars in clusters, and they
undergo significant mass loss upon evolving into HB stars.
Consequently, the progenitors of HB stars should be heavily mass
segregated.  HB stars themselves, however, have relatively low-masses
so that two-body relaxation and 
strong dynamical encounters should act to re-distribute them to wider
orbits within the cluster potential.  The HB 
lifetime is roughly constant at $10^8$ years \citep{iben91} and it is
comparable to or exceeds the core relaxation times for most of the
low-mass clusters in our sample \citep{harris96}.  Therefore, we might
expect the HB populations in these 
clusters to exhibit more extended radial profiles relative to 
more massive clusters.  This
would contribute to a sub-linear relationship between the numbers of
HB stars and the core masses.  Our uncertainties are sufficiently
large that this possibility cannot be entirely ruled out, however our
results are consistent with a linear relationship between HB numbers
and the core masses.  

\subsection{Additional Considerations} \label{final}

Recent observations have revealed the presence of
multiple stellar populations in a number of MW GCs
\citep[e.g.][]{pancino03}.  The majority of these cases have
been reported in very massive clusters.  Moreover, their existence is
thought to be related to the chemical properties of GCs, in
particular an observed anticorrelation between their sodium and oxygen 
abundances.  In turn, these chemical signatures have been argued to be
linked to the cluster metallicity, mass and age \citep{carretta10}.  

We identified clusters in our sample currently known to host multiple stellar
populations, but none of these were clear outliers in
our plots.  Consequently, the effects had on our results by multiple stellar
populations remains unclear.  It is certainly possible that multiple
stellar populations have contributed to the uncertainties for
the weighted lines of best fit performed for the relations in 
Figure~\ref{fig:Mshell_vs_Npop_2x2_cum} and
Figure~\ref{fig:Mshell_vs_Npop_2x2_noncum}.  It is difficult to
quantify the possible severity of this effect, however, given the
limited evidence linking multiple stellar populations to cluster
properties.  

Although the uncertainties are sufficiently large that the slopes are
consistent with being linear at the $3-\sigma$ confidence level for
all stellar populations when 
performing the comparisons with the total stellar mass, the reported
slopes are typically less than unity within the $1-\sigma$, and often 
even the $2-\sigma$, confidence interval.  This does not appear 
to result from the fact that we have 
obtained our estimates for the total stellar masses 
by numerically integrating 3-dimensional density distributions and are
comparing to number counts, which are projected quantities.  To
address this, we also tried obtaining the total stellar masses by
numerically integrating 2-dimensional surface brightness profiles so
that we are consistently comparing only projected quantities.  Despite
this, our results remain unchanged and the new slopes agree with the 
old ones to within 
one standard deviation for all stellar populations.  Another
possibility to account for this trend that is perhaps worth 
considering is a systematic dependence of the mass-to-light
ratios of clusters on their total mass.  There are two ways this could
have affected our analysis.  First, stellar remnants have been shown to 
affect the dynamical evolution of clusters, and therefore the sizes of
their cores \citep[e.g.][]{lee91, trenti10}.  It follows that, if the
number of stellar remnants depends on the cluster mass, then this
could contribute to an additional underlying dependence of 
the core radius on the cluster mass.  This could perhaps
arise as a result of the fact that the ratio of the rate of stellar
evolution to the rate of dynamical evolution is larger in more massive
clusters, since in general the rate of two-body relaxation decreases with
increasing cluster mass whereas the rate of stellar evolution is
independent of the cluster mass.  Coupled with their deeper gravitational
potential wells, this could contribute to more massive clusters
retaining more stellar remnants.  Second, variations in the
mass-to-light ratios of clusters can also occur as a result of 
changes in the average stellar mass (not including stellar remnants)
\citep{kruijssen09}.  That is, we can approximate the total stellar
mass contained in the core as:
\begin{equation}
\label{eqn:M_core}
M_{core} \sim \frac{4}{3}{\pi}\frac{M}{L}\rho_0r_c^3 \sim mN_{core},
\end{equation}
where $M/L$ is the mass-to-light ratio, $\rho_0$ is the central 
luminosity density, $m$ is the average stellar mass and $N_{core}$
is the total number of stars in the core.  Based on our results,
$M_{core} \propto N_{core}^{0.9}$, where we have used the total number
of stars in the core with magnitudes brighter than 0.5 mag below the
MSTO as a proxy for $N_{core}$.  This could suggest that $m \propto
N_{core}^{-0.1} \propto M_{core}^{-0.1}$.  In other words, the
average stellar mass in the core decreases weakly with increasing core
mass.  This could in part be due to the fact that more massive clusters 
should be less dynamically evolved than their less massive
counterparts, and should therefore be less depleted of their low-mass
stars due to stellar evaporation induced by two-body relaxation
\citep[e.g.][]{ambartsumian38, spitzer58, henon60, demarchi10}.
Similarly, mass 
segregation should also tend to operate more rapidly in low-mass
clusters, which acts to migrate preferentially massive stars into the
core \citep{spitzer69, spitzer71, farouki82, shara95, king95,
  meylan97}.  
Alternatively, differences in the stellar mass function in the 
core could result from variations in the degree of primordial mass
segregation, or even variations in the initial stellar mass function.

We have assumed throughout our analysis that the core mass is a
suitable proxy for the total cluster mass.  We have checked that these
two quantities are 
indeed correlated, however this does not tell the whole story since we
are also using the total cluster mass as a proxy for the degree of
dynamical evolution.  The central concentration 
parameter, defined as the logarithm of the ratio of the tidal to core
radii, describes the degree to which a cluster is centrally
concentrated.  Previous studies have shown that there exists a weak
correlation between the concentration parameter and the total cluster
mass \citep[e.g.][]{djorgovski94, mclaughlin00}.  In order to better use our
technique to reliably probe the effects of the cluster
dynamics on the sizes and radial distributions of the different
stellar populations, the concentration parameter should ideally be
accounted for when applying our normalization technique in future
studies.  It is not yet clear how the concentration parameter can be
properly absorbed into the normalization, however its effect on our
analysis should be small given the weak dependence on cluster mass.  

The assumption that the degree of dynamical evolution experienced by a
given cluster depends only on its mass is also incorrect.  Two-body 
relaxation has been shown to dominate cluster evolution for a
significant fraction of the lives of old MW GCs
\citep[e.g.][]{gieles11}, however other effects can also play a
significant role.  For example, stellar evolution is known to affect
the dynamical evolution of star 
clusters, although its primary role is played during their early
evolutionary phases 
\citep[e.g.][]{applegate86, chernoff90, fukushige95}.  Tidal effects
from the Galaxy have also been shown to play an important role in
deciding the dynamical fates of clusters by increasing the rate of
mass loss across the 
tidal boundary \citep[e.g.][]{heggie03}.  Consequently, clusters with
small perigalacticon distances should appear more dynamically evolved
than their total mass alone would suggest.  This effect can 
be significant, and has likely contributed to increasing the uncertainties
found for the comparisons to the total stellar mass.  Therefore, tidal
effects from the Galaxy should also ideally 
be absorbed into our normalization technique in future studies.  This
can be done by using the perigalacticon distances of clusters as a
rough proxy for the degree to which tides from the Galaxy should have
affected their internal dynamical evolution \citep{gieles11}.  

Interestingly, tides could also
help to explain why the uncertainties for the comparisons to the total
number of stars in each circle are considerably smaller than for the
comparisons to the total stellar mass.  We have used 
the number of stars with magnitudes brighter than 0.5 mag below the
MSTO as a proxy for the total number of stars.  Consequently, we are
comparing stars within a very narrow mass range, so that all 
populations of interest should have been comparably affected by
two-body relaxation (except, perhaps, for HB stars) independent of
tidal effects from 
the Galaxy.  In other words, tides should affect all stars above the
MSTO more or less equally, and this is consistent with our results.
It is also worth mentioning here that our King models consider only a
single stellar 
mass.  This assumption is not strictly true and could also be
contributing to increasing the uncertainties found for the comparisons
to the total stellar masses.

An additional concern is that we do not know if the clusters in our
sample are currently in a phase of core contraction or expansion.  
This has a direct bearing on the recent history of the stellar density
in the core, and therefore the degree to which stars in the core
should have been affected by close dynamical interactions.  These
effects are independent of two-body relaxation and occur on a
time-scale that is typically much shorter than the half-mass relaxation
time \citep{heggie03}.  The effects could be significant
in clusters that were recently in a 
phase of core-collapse but have since rebounded back out of this 
highly concentrated state.  This could occur, for example, as a result
of binary 
formation induced by 3-body interactions combined with their subsequent
hardening via additional encounters \citep{hut83, heggie03}.  In general,
binaries play an important role in the dynamical evolution of
clusters, and could have affected our results in a number of ways.
This is a difficult issue to address even qualitatively 
given how little is currently known about the binary populations in
globular clusters.  Theoretical models suggest, however, that the
time-scale for core contraction is often longer than a Hubble time, and
that this evolutionary phase will only come to an end once the central
density becomes sufficiently high for hardening encounters involving
binaries to halt the process \citep{fregeau09}.  It
follows that the cores of most MW GCs are expected to currently be in
a phase of core contraction.  This process is ultimately driven by
two-body relaxation, so that our assumption that the total
cluster mass provides a suitable proxy for the degree of dynamical
evolution is still valid.

In summary, our results suggest that effects related to the cluster
dynamics do not significantly affect 
the relative sizes of the different stellar populations above the
MSTO.  This is the case for 
at least RGB, HB and MSTO stars.  BSs, on the other hand, show
evidence for a sub-linear dependence of population size on the total
stellar mass contained within the same radius.  Whether or not the
cluster dynamics is responsible for this sub-linearity is still not
clear.  Notwithstanding, our results have provided 
evidence that mass segregation is not the dominant cause for this
result, although it will be necessary to redo the comparison performed
in this study with a larger spatial coverage in order
to fully address this question.  Further insight into the origin of
the sub-linearity found for BSs will be provided by reliable
binary fractions for the clusters in our sample, which are forthcoming
(Sarajedini 2010, private communication).

\clearpage

\begin{table}
\begin{center}
\caption{Stellar Population Catalogue}
\tiny
\begin{tabular}{@{}|p{1cm}@{}|p{1.5cm}@{}|p{1.7cm}@{}|@{}c@{}|@{}c@{}|@{}c@{}|@{}c@{}|@{}c@{}|@{}c@{}|@{}c@{}|@{}c@{}|@{}c@{}|@{}c@{}|@{}c@{}|@{}c@{}|c@{}|c@{}|c@{}|c@{}|c@{}|c@{}|c@{}|c@{}|@{}} 
\hline
Cluster ID  &  Alternate ID  &  Core Radius (in arcmin)  & \multicolumn{4}{|c|}{N$_{BS}$}           & \multicolumn{4}{|c|}{N$_{HB}$}           & \multicolumn{4}{|c|}{N$_{RGB}$}          & \multicolumn{4}{|c|}{N$_{MSTO}$}         & \multicolumn{4}{|c|}{N$_{TOT}$}          \\
\hline
            &              &                           & $< r_c$ & $< 2r_c$ & $< 3r_c$ & $< 4r_c$ & $< r_c$ & $< 2r_c$ & $< 3r_c$ & $< 4r_c$ & $< r_c$ & $< 2r_c$ & $< 3r_c$ & $< 4r_c$ & $< r_c$ & $< 2r_c$ & $< 3r_c$ & $< 4r_c$ & $< r_c$ & $< 2r_c$ & $< 3r_c$ & $< 4r_c$ \\
\hline
 104 &    47 Tuc    & 0.36 & 62 & 100 & 120 & 128 & 172 & 344 & 486 & 615 &  397 &  944 & 1454 & 1798 & 2190 & 5004 & 7300 & 9080 &   4874 & 11430 & 16985 & 21183 \\
1261 &              & 0.35 & 56 &  79 &  95 & 104 &  73 & 170 & 216 & 250 &  241 &  481 &  664 &  755 & 1102 & 2268 & 2953 & 3369 &   2713 &  5576 &  7347 &  8429 \\
1851 &              & 0.09 & 34 &  58 &  74 &  90 &  33 & 107 & 161 & 213 &   93 &  223 &  307 &  385 &  178 &  692 & 1128 & 1524 &    417 &  1418 &  2421 &  3400 \\
2298 &              & 0.31 & 27 &  32 &  37 &  38 &  16 &  41 &  56 &  63 &   61 &  120 &  158 &  186 &  208 &  429 &  568 &  662 &    549 &  1117 &  1490 &  1753 \\
3201 &              & 1.30 & 40 &  -- &  -- &  -- &  43 &  -- &  -- &  -- &  160 &   -- &   -- &   -- &  635 &   -- &   -- &   -- &   1691 &    -- &    -- &    -- \\
4147 &              & 0.09 & 16 &  26 &  30 &  34 &   7 &  18 &  35 &  44 &   23 &   61 &   93 &  120 &   89 &  206 &  316 &  400 &    234 &   569 &   844 &  1064 \\
4590 &      M 68    & 0.58 & 29 &  59 &  -- &  -- &  33 &  66 &  -- &  -- &  152 &  269 &   -- &   -- &  480 &  977 &   -- &   -- &   1321 &  2623 &    -- &    -- \\
5024 &      M 53    & 0.35 & 57 & 103 & 133 & 149 & 114 & 235 & 333 & 387 &  293 &  704 & 1059 & 1260 & 1215 & 2864 & 4106 & 4891 &   3118 &  7504 & 10730 & 12827 \\
5139 & $\Omega$ Cen & 2.37 & 49 &  87 &  -- &  -- & 408 & 762 &  -- &  -- & 1441 & 2592 &   -- &   -- & 4643 & 8637 &   -- &   -- &  12652 & 23178 &    -- &    -- \\
5272 &       M 3    & 0.37 & 74 & 111 & 127 & 135 & 153 & 311 & 379 & 413 &  496 &  995 & 1277 & 1387 & 1909 & 3828 & 5052 & 5512 &   4971 & 10020 & 13195 & 14429 \\
5286 &              & 0.28 & 82 & 120 & 138 & 144 & 218 & 413 & 530 & 599 &  442 &  970 & 1308 & 1535 & 1723 & 3666 & 4983 & 5876 &   4016 &  8934 & 12448 & 14826 \\
5466 &              & 1.43 & 30 &  -- &  -- &  -- &  37 &  -- &  -- &  -- &  123 &   -- &   -- &   -- &  487 &   -- &   -- &   -- &   1276 &    -- &    -- &    -- \\
5904 &       M 5    & 0.44 & 37 &  57 &  64 &  68 &  97 & 212 & 291 & 338 &  233 &  516 &  729 &  885 &  997 & 2260 & 3190 & 3843 &   2483 &  5700 &  8123 &  9846 \\
5927 &              & 0.42 & 28 &  71 &  93 & 122 &  91 & 207 & 294 & 358 &  188 &  513 &  748 &  922 & 1214 & 3043 & 4528 & 5667 &   2619 &  6714 & 10108 & 12688 \\
5986 &              & 0.47 & 57 &  88 &  -- &  -- & 220 & 386 &  -- &  -- &  614 & 1136 &   -- &   -- & 2359 & 4549 &   -- &   -- &   5756 & 11255 &    -- &    -- \\
6093 &      M 80    & 0.15 & 79 & 114 & 133 & 135 &  94 & 199 & 269 & 331 &  252 &  543 &  773 &  984 & 1045 & 2176 & 3090 & 3790 &   2008 &  4627 &  6840 &  8637 \\
6101 &              & 0.97 & 26 &  -- &  -- &  -- &  68 &  -- &  -- &  -- &  173 &   -- &   -- &   -- &  681 &   -- &   -- &   -- &   1798 &    -- &    -- &    -- \\
6121 &       M 4    & 1.16 & 11 &  18 &  -- &  -- &  21 &  46 &  -- &  -- &   52 &  126 &   -- &   -- &  243 &  574 &   -- &   -- &    553 &  1350 &    -- &    -- \\
6171 &     M 107    & 0.56 & 19 &  43 &  54 &  -- &  16 &  37 &  56 &  -- &   63 &  153 &  223 &   -- &  264 &  667 &  933 &   -- &    677 &  1688 &  2414 &    -- \\
6205 &      M 13    & 0.62 & 41 &  58 &  -- &  -- & 207 & 416 &  -- &  -- &  527 & 1162 &   -- &   -- & 1960 & 4250 &   -- &   -- &   5015 & 10973 &    -- &    -- \\
6218 &      M 12    & 0.79 & 28 &  50 &  -- &  -- &  32 &  68 &  -- &  -- &  114 &  245 &   -- &   -- &  447 & 1118 &   -- &   -- &   1127 &  2680 &    -- &    -- \\
6254 &      M 10    & 0.77 & 36 &  52 &  -- &  -- &  93 & 169 &  -- &  -- &  257 &  540 &   -- &   -- &  955 & 1985 &   -- &   -- &   2483 &  5165 &    -- &    -- \\
6304 &              & 0.21 & 19 &  36 &  51 &  67 &  27 &  65 &  95 & 112 &   82 &  207 &  313 &  397 &  453 & 1112 & 1657 & 2143 &    994 &  2584 &  3864 &  5008 \\
6341 &      M 92    & 0.26 & 41 &  73 &  84 &  91 &  60 & 126 & 177 & 217 &  140 &  367 &  540 &  684 &  543 & 1290 & 1896 & 2376 &   1409 &  3341 &  4943 &  6252 \\
6362 &              & 1.13 & 35 &  -- &  -- &  -- &  61 &  -- &  -- &  -- &  165 &   -- &   -- &   -- &  716 &   -- &   -- &   -- &   1844 &    -- &    -- &    -- \\
6535 &              & 0.36 &  7 &  11 &  12 &  12 &   8 &  18 &  24 &  26 &   21 &   42 &   56 &   72 &   54 &  107 &  174 &  227 &    165 &   338 &   493 &   629 \\
6584 &              & 0.26 & 36 &  54 &  63 &  -- &  52 &  95 & 135 &  -- &  217 &  386 &  482 &   -- &  788 & 1499 & 1863 &   -- &   2023 &  3810 &  4830 &    -- \\
6637 &      M 69    & 0.33 & 50 &  85 &  96 & 106 &  80 & 148 & 204 & 239 &  200 &  443 &  592 &  702 & 1067 & 2257 & 3063 & 3605 &   2413 &  5209 &  7129 &  8414 \\
6652 &              & 0.10 & 16 &  19 &  24 &  27 &  10 &  21 &  34 &  40 &   32 &   61 &   87 &  122 &  127 &  272 &  417 &  536 &    286 &   619 &   919 &  1218 \\
6723 &              & 0.83 & 39 &  -- &  -- &  -- & 113 &  -- &  -- &  -- &  354 &   -- &   -- &   -- & 1594 &   -- &   -- &   -- &   3777 &    -- &    -- &    -- \\
6779 &      M 56    & 0.44 & 21 &  41 &  48 &  49 &  44 &  99 & 133 & 158 &  128 &  302 &  435 &  528 &  411 &  993 & 1495 & 1875 &   1126 &  2679 &  3982 &  4912 \\
6838 &      M 71    & 0.63 & 17 &  45 &  -- &  -- &  10 &  30 &  -- &  -- &   36 &   95 &   -- &   -- &  144 &  385 &   -- &   -- &    355 &   960 &    -- &    -- \\
6934 &              & 0.22 & 35 &  54 &  57 &  60 &  50 & 100 & 137 & 163 &  150 &  322 &  431 &  508 &  612 & 1240 & 1681 & 1974 &   1528 &  3208 &  4308 &  5088 \\
6981 &      M 72    & 0.46 & 31 &  49 &  56 &  -- &  52 &  78 & 103 &  -- &  140 &  285 &  354 &   -- &  652 & 1272 & 1596 &   -- &   1594 &  3159 &  4000 &    -- \\
7089 &       M 2    & 0.32 & 83 & 129 & 143 & 150 & 277 & 551 & 729 & 838 &  535 & 1205 & 1652 & 1960 & 2264 & 4832 & 6603 & 7795 &   5394 & 12038 & 16669 & 19851 \\
\hline
\end{tabular}
\label{table:catalogue}
\end{center}
\end{table}

\clearpage

\begin{table}
\small
\caption{Lines of Best Fit for log(M$_{circle}$/10$^3$) Versus
  log(N$_{pop}$) \label{table:Mshell_vs_Npop_cum}}
\begin{tabular}{|l|p{3cm}|p{3cm}|p{3cm}|p{3cm}|}
\hline
Circle      &        BS        &        RGB         &         HB         &       MSTO      \\
\hline
$<$ r$_c$   & log(N$_{BS}$) = (0.39 $\pm$ 0.05)log(M$_{< r_c}$/10$^3$) + (1.22 $\pm$ 0.05)  &    log(N$_{RGB}$) = (0.95 $\pm$ 0.11)log(M$_{< r_c}$/10$^3$) + (1.36 $\pm$ 0.11)    & log(N$_{HB}$) = (0.95 $\pm$ 0.06)log(M$_{< r_c}$/10$^3$) + (0.92 $\pm$ 0.06)   & log(N$_{MSTO}$) = (0.90 $\pm$ 0.07)log(M$_{< r_c}$/10$^3$) + (2.03 $\pm$ 0.07)    \\
$<$ 2r$_c$  & log(N$_{BS}$) = (0.36 $\pm$ 0.05)log(M$_{< 2r_c}$/10$^3$) + (1.26 $\pm$ 0.08)  &    log(N$_{RGB}$) = (0.87 $\pm$ 0.08)log(M$_{< 2r_c}$/10$^3$) + (1.27 $\pm$ 0.12)    & log(N$_{HB}$) = (0.85 $\pm$ 0.06)log(M$_{< 2r_c}$/10$^3$) + (0.84 $\pm$ 0.09)   & log(N$_{MSTO}$) = (0.82 $\pm$ 0.06)log(M$_{< 2r_c}$/10$^3$) + (1.98 $\pm$ 0.09)    \\
$<$ 3r$_c$  & log(N$_{BS}$) = (0.47 $\pm$ 0.04)log(M$_{< 3r_c}$/10$^3$) + (1.02 $\pm$ 0.08)  &    log(N$_{RGB}$) = (0.80 $\pm$ 0.06)log(M$_{< 3r_c}$/10$^3$) + (1.26 $\pm$ 0.12)    & log(N$_{HB}$) = (0.79 $\pm$ 0.08)log(M$_{< 3r_c}$/10$^3$) + (0.83 $\pm$ 0.14)   & log(N$_{MSTO}$) = (0.82 $\pm$ 0.11)log(M$_{< 3r_c}$/10$^3$) + (1.86 $\pm$ 0.20)    \\
$<$ 4r$_c$  & log(N$_{BS}$) = (0.45 $\pm$ 0.05)log(M$_{< 4r_c}$/10$^3$) + (1.01 $\pm$ 0.12)  &    log(N$_{RGB}$) = (0.75 $\pm$ 0.07)log(M$_{< 4r_c}$/10$^3$) + (1.28 $\pm$ 0.15)    & log(N$_{HB}$) = (0.75 $\pm$ 0.09)log(M$_{< 4r_c}$/10$^3$) + (0.83 $\pm$ 0.19)   & log(N$_{MSTO}$) = (0.78 $\pm$ 0.12)log(M$_{< 4r_c}$/10$^3$) + (1.86 $\pm$ 0.25)    \\
\hline
\end{tabular}
\end{table}

\begin{table}
\small
\caption{Lines of Best Fit for log(M$_{annulus}$/10$^3$) Versus
  log(N$_{pop}$) \label{table:Mshell_vs_Npop_noncum}}
\begin{tabular}{|l|p{3.5cm}|p{3.5cm}|p{3.5cm}|p{3.5cm}|}
\hline
Annulus                 &         BS        &        RGB        &        HB       &       MSTO      \\
\hline
 r$_c$ $<$ r $<$ 2r$_c$ & log(N$_{BS}$) = (0.27 $\pm$ 0.08)log(M$_{r_c < r < 2r_c}$/10$^3$) + (1.04 $\pm$ 0.13)  &    log(N$_{RGB}$) = (0.80 $\pm$ 0.06)log(M$_{r_c < r < 2r_c}$/10$^3$) + (1.22 $\pm$ 0.08)    & log(N$_{HB}$) = (0.77 $\pm$ 0.06)log(M$_{r_c < r < 2r_c}$/10$^3$) + (0.79 $\pm$ 0.09)   & log(N$_{MSTO}$) = (0.76 $\pm$ 0.05)log(M$_{r_c < r < 2r_c}$/10$^3$) + (1.91 $\pm$ 0.08)    \\
2r$_c$ $<$ r $<$ 3r$_c$ & log(N$_{BS}$) = (0.39 $\pm$ 0.09)log(M$_{2r_c < r < 3r_c}$/10$^3$) + (0.52 $\pm$ 0.16)  &    log(N$_{RGB}$) = (0.79 $\pm$ 0.11)log(M$_{2r_c < r < 3r_c}$/10$^3$) + (0.97 $\pm$ 0.17)    & log(N$_{HB}$) = (0.68 $\pm$ 0.10)log(M$_{2r_c < r < 3r_c}$/10$^3$) + (0.67 $\pm$ 0.16) & log(N$_{MSTO}$) = (0.78 $\pm$ 0.12)log(M$_{2r_c < r < 3r_c}$/10$^3$) + (1.64 $\pm$ 0.20)    \\
3r$_c$ $<$ r $<$ 4r$_c$ & log(N$_{BS}$) = (0.15 $\pm$ 0.21)log(M$_{3r_c < r < 4r_c}$/10$^3$) + (0.77 $\pm$ 0.35)  &    log(N$_{RGB}$) = (0.63 $\pm$ 0.11)log(M$_{3r_c < r < 4r_c}$/10$^3$) + (1.04 $\pm$ 0.18)    & log(N$_{HB}$) = (0.69 $\pm$ 0.18)log(M$_{3r_c < r < 4r_c}$/10$^3$) + (0.45 $\pm$ 0.31) & log(N$_{MSTO}$) = (0.69 $\pm$ 0.16)log(M$_{3r_c < r < 4r_c}$/10$^3$) + (1.60 $\pm$ 0.27)    \\
\hline
\end{tabular}
\end{table}

\clearpage

\appendix

\section{Stellar Population Selection Criteria} \label{appendix}

In this section, we present our selection criteria for BS, RGB, HB and
MSTO stars.  Our method is similar to that described in \citet{leigh07},
and we have used this as a basis for the selection criteria presented
in this paper.  First, we define a location for the MSTO in the
(F606W-F814W)-F814W plane using our isochrone fits.  The MSTO is
chosen to be the bluest point along the MS of each isochrone, which we
denote by ((V-I)$_{MSTO}$,I$_{MSTO}$).  In order to distinguish BSs
from MSTO stars, we impose the conditions:

\begin{equation}
\label{eqn:bs_msto1}
F814W \le m_1(F606W-F814W) + b_{11},
\end{equation}
where the slope of this line is $m_1 = -9$ and its y-intercept is
given by: 
\begin{equation}
\label{eqn:b_11}
b_{11} = (I_{MSTO}-0.10)-m_1((V-I)_{MSTO}-0.10)
\end{equation}

Similarly, we distinguish BSs from HB stars by defining the following
additional boundaries:
\begin{eqnarray}
\label{bs_hb1}
F814W &\ge& m_1(F606W-F814W) + b_{12} \\
F814W &\ge& m_2(F606W-F814W) + b_{21} \\
F814W &\le& m_2(F606W-F814W) + b_{22} \\
F814W &\ge& m_{HB}(F606W-F814W) + b_{HB} \\
(F606W-F814W) &\ge& (V-I)_{HB} \\
F814W &\le& I_{MSTO},
\end{eqnarray}
where $m_2 = 6$, $m_{HB} = -1.5$ and $(V-I)_{HB} = (V-I)_{MSTO} -
0.4$.  We also define:
\begin{eqnarray}
\label{b_12}
b_{12} &=& (I_{MSTO}-0.55)-m_1((V-I)_{MSTO}-0.55) \\
b_{21} &=& (I_{MSTO}-0.80)-m_2((V-I)_{MSTO}+0.10) \\
b_{22} &=& (I_{MSTO}+0.30)-m_2((V-I)_{MSTO}-0.20),
\end{eqnarray}
and $b_{HB} = I_{HB} + 1.2$, where $I_{HB}$ roughly corresponds to the 
mid-point of points that populate the HB and is chosen
by eye for each cluster so that our selection criteria best fits the
HB in all of the CMDs in our sample.   

We apply a similar set of conditions to the RGB in order to select
stars belonging to this stellar population.  These boundary conditions
are:
\begin{eqnarray}
\label{rgb_1}
F814W &\ge& m_{HB}(F606W-F814W) + b_{HB} \\
F814W &\ge& m_{RGB}(F606W-F814W) + b_{31} \\
F814W &\le& m_{RGB}(F606W-F814W) + b_{32} \\
F814W &\le& I_{RGB},
\end{eqnarray}
where $m_{RGB} = -23$, $I_{RGB}$ is defined as the F814W magnitude
corresponding to a core helium mass of 0.08 M$_{\odot}$ and:
\begin{eqnarray}
\label{b_31_and_b_32}
b_{31} &=& (I_{MSTO}-0.60)-m_{RGB}((V-I)_{MSTO}+0.05) \\
b_{32} &=& (I_{MSTO}-0.60)-m_{RGB}((V-I)_{MSTO}+0.25)
\end{eqnarray}

Core helium-burning stars, which we refer to as HB stars, are selected
if they satisfy one of the following sets of criteria:
\begin{eqnarray}
\label{hb1}
F814W &\ge& m_{HB}(F606W-F814W) + (b_{HB}-1.0) \\
F814W &\le& m_{HB}(F606W-F814W) + b_{HB} \\
(F606W-F814W) &\le& (V-I)_{MSTO} + (V-I)_{HB},
\end{eqnarray}
\begin{eqnarray}
\label{hb2}
F814W &>& m_{HB}(F606W-F814W) + b_{HB} \\
F814W &\le& I_{MSTO} + 2.5 \\
(F606W-F814W) &<& (V-I)_{MSTO} - 0.4,
\end{eqnarray}
or
\begin{eqnarray}
\label{hb3}
F814W &<& m_1(F606W-F814W) + b_{12} \\
F814W &>& m_{HB}(F606W-F814W) + b_{HB} \\
(F606W-F814W) &\ge& (V-I)_{MSTO} - 0.4
\end{eqnarray}
We define $(V-I)_{HB}$ on a cluster-by-cluster basis in order to
ensure that we do not over- or under-count the number of HB stars.
This is because the precise value of (F606W-F814W) at which the HB
becomes the RGB varies from cluster-to-cluster.  In addition, the
precise location of the transition in the cluster CMD between HB 
and EHB stars remains poorly understood.  To avoid this ambiguity, we
consider HB and EHB stars 
together throughout our analysis, and collectively refer to all core
helium-burning stars as HB stars throughout this paper.

Finally, MSTO stars are selected according to the following criteria:
\begin{eqnarray}
\label{msto_1}
F814W &>& I_{RGB} \\
F814W &>& m_1(F606W-F814W) + b_{11} \\
F814W &\le& (V-I)_{MSTO}
\end{eqnarray}

\section*{Acknowledgments}

We would like to thank Ata Sarajedini, Aaron Dotter and Roger Cohen
for providing the data on which this study was based and for their
extensive support in its analysis.  We would also like to thank Evert
Glebbeek for useful discussions.  This research has been
supported by NSERC and OGS.

\bsp

\label{lastpage}

\end{document}